\documentclass[pra,showpacs,showkeys,nofootinbib]{revtex4}

\usepackage{amsmath,amssymb}
\usepackage{epsfig}
\usepackage{color}
\newcommand{\n}{\nonumber}
\newcommand{\be}{\begin{equation}}
\newcommand{\ee}{\end{equation}}
\newcommand{\bea}{\begin{eqnarray}}
\newcommand{\eea}{\end{eqnarray}}

\begin{document}

\title{Convection-Diffusion-Reaction equation with similarity solutions}
\author{Choon-Lin Ho and Chih-Min Yang}%\\
%\addresss
\affiliation{Department of Physics, Tamkang University,
Tamsui 25137, Taiwan}

%\date{May 30, 2018}

%\maketitle  % for LaTex

\begin{abstract}

We consider similarity solutions of the generalized convection-diffusion-reaction
equation with both space- and time-dependent convection, diffusion and reaction terms. By  introducing  the
similarity variable, the reaction-diffusion equation is reduced to an
ordinary differential equation. Matching the
resulting ordinary differential equation with known exactly solvable equations, one can obtain corresponding exactly solvable convection-diffusion-reaction systems.   Some representative examples of exactly solvable systems are presented.
We also describe how an equivalent convection-diffusion-reaction system can be constructed which admits the same similarity solution of another convection-diffusion-reaction system. 

\end{abstract}

% 05.10.Gg  Stochastic analysis methods (Fokker-Planck, Langevin, etc.)
%05.90.+m   Other topics in statistical physics, thermodynamics, and nonlinear dynamical systems (restricted to new topics in section 05)
%02.50.Ey   Stochastic processes


\maketitle
                          
%\newpage

%%%%%%%%%%%%%%%%%%%%%%
\section{Introduction}

The Convection-Diffusion-Reaction (CDR) equation is a class of second order differential that is widely employed to model phenomena that involve the change of concentration/population of one or more substances/species distributed in space  under the influence of three processes: local reaction which modify the concentration/population, diffusion which causes the substances/species  to spread in space, and convection/drifting under the influence of external forces \cite{GK1,GK2,HM,CP}.   This includes as special cases the Fokker-Planck equation [5-13] and the reaction-diffusion equation [14-25]. The CDR equation has found many important applications in physics, chemistry, astrophysics, engineering, and biology. 

In view of its broad applicability, it is thus desirable to obtain analytic solutions of the CDR equation for as many systems as possible. However,
just as any equation in sciences, solving the CDR equation exactly is in general a formidable task, except in a few simplified cases.   Interestingly, many analytic solutions were found in the form of travelling wave solutions \cite{GK1,GK2,HM,CP}. 

Different forms of analytic solutions may be possible for the CDR equation.  In our previous work, we have studied exact  solvability of the Fokker-Planck \cite{Ho1,Ho2,Ho3} and reaction-diffusion equations \cite{HL} in terms of similarity solutions.
Here we would like to extend our previous consideration to the CDR equation. 

One advantage of the similarity method is that it allows one to
reduce the partial differential equation under consideration to an
ordinary differential equation which is generally easier to solve, provided that the original equation
possesses proper scaling property under certain scaling
transformation of the basic variables.   

We shall first  discuss the
scaling properties of the CDR equation. This includes the scaling forms of the relevant function, and  the corresponding similarity variable.  The equation of continuity is then considered, which is used to identify two types of  scaling behaviours of the CDR equation, namely, systems in which particle number is conserved or not. 
Some examples of these two types of scaling CDR systems are presented.  Finally, we briefly discuss how an equivalent CDR system can be constructed which admits the same similarity solution of another CDR system. 

%%%%%%%%%%%%%%%%%%%
\section{Scaling form of CDR equation and similarity variable}

The $(1+1)$-dimension CDR equation to be studied in this paper is taken to have the following general form
\bea
&&\frac{\partial W(x,t)}{\partial t}=\frac{\partial}{\partial x}\left(D(W,x,t)\frac{\partial}{\partial x} W(x,t)\right) \n\\
&&~~~~-\frac{\partial}{\partial x}\left(C(W,x,t)\ W(x,t)\right)+ R(W,x,t),
\label{CDR}
\eea
where $W(x,t)$ is the particle number function,
$D(W,x,t), C(W,x,t) $  and $R(W,x,t)$ are the diffusion, the convection, and  the
reaction term, respectively.  We use the term  ``particle" to denote generally the number of basic member of a substance or a specie.  The domains we shall consider in this paper are the real line
$x\in (-\infty,\infty)$, or  the half lines $x\in [0,\infty)$.  Cases with finite domains, which correspond to systems with moving boundaries, can be considered similarly \cite{Ho2}.  We leave the possibility that $C, D$ and $R$ could be functions of $W$
(when no confusion arises, we shall often omit the independent variables of a function for simplicity and clarity of presentation).

If the CDR possesses scaling symmetry, then similarity solution is possible.
Suppose under the scale transformation
\be
x=\epsilon^a \,\bar{x}\;\;\;,\;\;\; t=\epsilon^b \,\bar{t},
\ee
  all the relevant functions of the CDR equation scale as
\bea
&&W(x,t)=\epsilon^\omega \bar{W}(\bar{x},\bar{t}),~~
C(W,x,t) =\epsilon^c \bar{C}(\bar{W}, \bar{x},\bar{t});\;\;\label{scale}\\
&&D(W,x,t) =\epsilon^d \bar{D}(\bar{W}, \bar{x},\bar{t}),\n\;\;
R(W,x,t)=\epsilon^r \bar{R}(\bar{W}, \bar{x},\bar{t}).
\eea
Here the scale factor $\epsilon$ and the scaling exponents $a, b, c, d, r$ and $\omega$ are real parameters.

CDR  equation possesses scaling symmetry under the above transformation  if  the form of the CDR equation expressed in terms of the transformed quantities is the same as the form of the original equation.   This is the case if the exponents are related by
 $b=2a-d=a-c=\omega-r$.

Under this situation, one can transform the second order CDR equation into an
ordinary differential equation by introducing the similarity variable $z$, which can be defined as
\be
z\equiv\frac{x}{t^{\alpha}}, ~~\mbox{where}~
\alpha=\frac{a}{b}\;\;\;,\; b\neq 0\;.
\ee
Suppose the  functions $C, D, R$ and $W$ have the following scaling forms:
\bea
W(x,t)&=&t^\mu y(z), ~~C(W,x,t)=t^\gamma\tau(z),\n\\
~~ D(W,x,t)&=&t^\delta \sigma(z), ~~ R(W,x,t)=t^\rho \rho(z).
\eea
From Eq. (\ref{scale}) together with the scaling conditions $b=2a-d=a-c=\omega-r$,  one has
\be
\mu=\frac{\omega}{b}, ~~ \gamma=\frac{c}{b}=\alpha-1,~~ \delta=\frac{d}{b}=2\alpha-1,~~\rho=\frac{r}{b}=\mu-1.
\ee
Thus $\alpha$ and $\mu$ are the only two independent scaling exponents of the CDR equation.

In terms of these scaling forms, Eq.~(\ref{CDR}) reduces to an ordinary differential equation
\be
\frac{d}{dz}\left(\sigma\frac{d}{dz}y\right) + \alpha z\frac{dy}{dz}-\frac{d}{dz}\left(\tau y\right) -\mu y+ \rho(z)=0,
\label{ODE}
\ee
or equivalently,
\be
\sigma y''+(\sigma' +\alpha\,z-\tau )\,y' -(\tau'+\mu)\, y+\rho=0.
\label{ODE2}
\ee
Here ``prime" represents derivative with respect to $z$.
Note  that when $\mu=-\alpha$, which we will encounter below, Eq.\,(\ref{ODE}) reduces to
\be
\frac{d}{dz}\left(\sigma\frac{d}{dz}y + \left(\alpha z -\tau\right)y\right)+ \rho(z)=0.
\label{CDR2}
\ee

Next we shall consider the condition imposed by
the continuity in the change of the particle number of the system,
i.e., the equation of continuity.

%%%%%%%%%%%%%%%%%
\section{Equation of continuity}

The total number $N$ of the system is related to the density function $W(x,t)$ by
\be
N=\int_{\cal D}\,W(x,t)\,dx=t^{\alpha+\mu}\int_{\cal D}\,
y(z)\,dz,
\label{N}
\ee
where $\cal D$ is the domain of the independent variable.  For simplicity, we use the same notation $\cal{D}$ for both the variable $x$, and the corresponding similarity variable $z$. 

 Eq.\,(\ref{N}) distinguishes two different situations: $\alpha+\mu\neq 0$ and $\alpha+\mu=0$.
It is obvious from this equation that $N$ is conserved if and only if  $\mu=-\alpha$.

Eq.\,(\ref{N}) implies that
\be
\frac{dN}{dt}=(\alpha+\mu)t^{\alpha+\mu-1} \left(\int_{\cal D}\,
y(z)\,dz\right).
\label{dN}
\ee
On the other hand, from Eq.\,(\ref{CDR}) one has
\bea	
\frac{dN}{dt}=t^{\alpha+\mu-1} \Delta \left(\sigma\frac{dy}{dz}-\tau y\right)_{\partial {\cal D}}+ t^{\alpha+\rho}  \int_{\cal D} \rho(z)\,dz.
\eea
Here $\partial{\cal D}$ denotes the boundaries of the domain $\cal D$, and $\Delta(\cdots)_{\partial {\cal D}}$ the difference of the terms in the bracket at the boundaries.

In view of $\rho=\mu-1$, one has
\be
 (\alpha+\mu)\int_{\cal D}\,
y(z)\,dz= \int_{\cal D}\, \rho(z)\,dz + \Delta \left(\sigma\frac{dy}{dz}-\tau y\right)_{\partial {\cal D}}.
\label{EOCa}
 \ee
It is interesting to note that, by integrating Eq.\,(\ref{ODE}) and replacing the integral of $\rho(z)$ in Eq.\,(\ref{EOCa}), the equation of continuity simplifies to
\be
\Delta( \alpha z y)=0.
\label{EOCa2}
\ee

Below  we discuss separately the cases for $\mu=-\alpha$ and $\mu\neq -\alpha$.

%%%%%%%%%%%%%%%%%%%%%
\section{Cases with $\mu= -\alpha$}

In this case $y(z)$ satisfies eq.\,(\ref{CDR2}). 
As $N$ is conserved when  $\mu=-\alpha$, one can normalize $W(x,t)=t^{-\alpha}y(z)$  and treat it as the probability distribution function.  Thus the number of particle is
\be
N=\int_{\cal D}\,
y(z)\,dz,
\label{N2}
\ee
and the equation of continuity (\ref{EOCa}) becomes
\be
\int_{\cal D}\, \rho(z)\,dz + \Delta \left(\sigma\frac{dy}{dz}-\tau y\right)_{\partial {\cal D}}=0.
\label{EOCb}
 \ee
 
 Any choice of the set of functions $\sigma(z), \rho(z)$ and $\tau(z))$ such that $y(z)$ is normalizable and Eq.(\ref{EOCb}) is satisfied defines an exactly solvable CDR system with similarity solution.  Formally the solution $y(z)$ is
 \bea
 y(z)&=&-e^{-\int^z\,dz\,
\frac{\alpha z-\tau}{\sigma}}\label{GS}\\
&&~~~\times \left(\int^z\,dz e^{\int^z
\frac{\alpha z-\tau}{\sigma}dz}\frac{1}{\sigma}\, \int^z \,dz\,\rho(z)+C\right),
\n
\eea 
where $C$ is an integration constant. 
 
For $y(z), \sigma(z)$ and $\tau(z)$ such that $\Delta (\sigma\, y' - \tau y)_{\partial {\cal D}}=0$,  one has $\int_{\cal D}\, \rho (z) dz=0$.   This is most easily satisfied if $\rho(z)$ is a total differential, i.e., $\rho(z)=-d{\bar\rho}(z)/dz$ for some function $\bar{\rho}(z)$ which vanishes at the boundaries, or  any function $\rho(z)$ that is anti-symmetric w.r.t. the mid-point of the domain $\cal{D}$ in the similarity variable $z$.   This latter situation is possible only if  $\cal{D}$  is the whole line or a finite domains in the $z$-space  (corresponding to moving boundaries in the $x$-space), and is not possible for the half-line. 
  
  If $\rho(z)$ is anti-symmetric w.r.t. the midpoint of $\cal D$, the Eq.\,(\ref{CDR2}) implies that $\sigma(z)$ is symmetric, while $\tau(z)$ and $y(z)$ are anti-symmetric. But an anti-symmetric $y(x)$ gives $N=0$ from Eq.\,(\ref{N2}). Thus this case is not of interest.
  
So we take $\rho(z)$ to be a total differential.
 This implies that Eq.\,(\ref{CDR2}) is a total derivative, and can be integrated once to give
 \be
\sigma y'+ \left(\alpha z -\tau\right)y-{\bar\rho}={\rm constant}.
\label{ODE3}
\ee
 
If the functions on the l.h.s. of Eq.\,(\ref{ODE3}) vanish at the boundaries,  the constant equals zero and we need only to consider the following equation instead
  \be
\sigma y'+ \left(\alpha z -\tau\right)y-{\bar\rho}=0.
\label{ODE4}
\ee

There are two situations that one can consider : Fokker-Planck type and non-Fokker-Planck type. To construct exactly solvable CDR systems, one can follow exactly the procedure presented in \cite{HL}.   So we shall give only the main ideas here.

% ------    FK type --------
\subsubsection*{$\bullet$~{\bf Fokker-Planck type}}

Let the function ${\bar\rho}(z)$  be proportional to $y(z)$, i.e.
${\bar\rho}(z)=\beta(z)y(z)$ for some function $\beta(z)$. In this case the CDR  is of the Fokker-Planck type, where the function $\beta(z)+\tau(z)$ plays the role of the drift coefficient.   Integrating Eq.~(\ref{ODE4}) once gives
\be
y(z)\propto \exp\left(\int^z\,dz \, \frac{\beta(z)+\tau(z)-\alpha z}{\sigma(z)}\right).
\label{FK}
\ee
Hence for any choice of $\sigma(z), \beta(z)$ and $\tau(z)$ are such that $y(z)$ in Eq.\,(\ref{FK}) is integrable and normalizable, one has an exactly solvable CDR system. This is exactly the same as the way to obtain similarity solutions of the Fokker-Planck equations discussed in \cite{Ho1,Ho2,Ho3}.   All the cases presented there for the Fokker-Planck equations  can be carried over to this type of CDR equations.  This includes cases related to solutions with moving boundaries, and solutions involving the recently discovered exceptional orthogonal polynomials, can be found in Ref.\,\cite {Ho2} and \cite{Ho3}, respectively.

%-------     Non FK type --------
\subsubsection*{$\bullet$~{\bf Non-Fokker-Planck type}}

For ${\bar\rho}(z)$ not proportional to $y(z)$, i.e., ${\bar\rho}(z)\neq \beta(z)y(z)$, 
the general solution of Eq.\,(\ref{GS}) is
\be
 y(z)=e^{-\int^z
\frac{\alpha z-\tau}{\sigma}dz}\left(\int^z e^{\int^z
\frac{\alpha z-\tau}{\sigma}dz}\frac{\bar\rho}{\sigma}dz +C\right), 
\label{NFK}
\ee
where $C$ is a constant of  integration. Any choice of $\sigma(z), \tau(z)$ and $\bar\rho(z)$ such that $y(z)$ is exactly integrable  and $W(x,t)$ is nomalizable furnishes a solvable RD system.

Again, the procedures presented in \cite{HL} can be carried over to construct exactly solvable CDR systems of this type.

%%%%%%%%%%%%%%%%%%%%%
\section{Cases with $\mu\neq -\alpha$}

When $\mu\neq -\alpha$, the number of particles does not conserve. 

Formally one can construct a solvable CDR system as follows.
Defining $\tilde{\rho}(z)\equiv (\tau'+\mu )y(z)-\rho(z)$, we rewrite Eq.~(\ref{ODE2}) as
 \be
 y'' +\frac{\sigma' +\alpha\,z-\tau}{\sigma} y' =\frac{\tilde\rho}{\sigma}.
 \label{ODE5}
 \ee
By first treating this equation as a first order differential equation for $y^\prime$ and using the method of integrating factor, one obtains the formal solution of Eq.~(\ref{ODE5}) as
\bea
y(z)&=&\int^z\,dz'\, e^{-\gamma(z')} \left[\int^{z'} \,dz''\,  e^{\gamma(z'')}\frac{\tilde{\rho}(z'')}{\sigma(z'')}+C^\prime\right]+C,\n\\
&& ~~~\gamma (z)\equiv \int^z\,dz\,\frac{\sigma'(z) +\alpha\,z-\tau(z)}{\sigma(z)},
\label{y-int}
 \eea
 where $C$ and $C^\prime$ are integrating constants. 
 One then looks for the functions $\sigma(z)$ and  $\tilde{\rho}(z)$ that make Eq.~(\ref{y-int}) integrable. An exactly solvable CDR system is then  given by the functions $\sigma(z)$, $y(z)$ and $\rho(z)=(\tau'(z)+\mu )y(z)-\tilde{\rho}(z)$.

Of course, while the above recipe gives a formal way to find exactly solvable CDR systems, it is in general not easy to determine the required functions 
$\sigma(z)$ and  $\tilde{\rho}(z)$.   A more practical way to look for exactly solvable CDR systems is to match Eq.~(\ref{ODE5}) with the known exactly solvable 2nd-order differential equations.  In what follows we shall present some examples based on this latter method.

%----------- Ex. 1 -----
\subsubsection*{$\bullet$~{\bf Example 1 : $ \sigma=1, \tau=\alpha z$ }}
\medskip

Taking $\sigma=1, \tau=\alpha z$, Eq.\,(\ref{ODE2}) becomes
\[
y''-(\alpha+\mu)\,y +\rho=0, ~~z\in (-\infty, \infty).
\]
By matching this equation with the differential equation in $\S 2.1.2.3$ of \cite{PZ}, 
\[y'' -(c^2 z^2 -c)\,y=0,~~c>0,
\]
which admits  a particular solution 
 \be
 y_0=e^{-\frac12 c z^2},
 \ee
 we get
 \[
 \rho=\left(-c^2 z^2 +c+\alpha+\mu\right)\,y.
 \]
 
we get a CDR system as follows:
\bea
C(x,t)&=&\alpha\, \frac{x}{t},\n\\
D(x,t)&=&t^{2\alpha-1},~~y(z)=e^{-\frac12 c z^2},~~x\in (-\infty, \infty),~\n\\
R(x,t)&=&t^{\mu-1}\left[-c^2\left(\frac{x}{t^\alpha}\right)^2+c+\alpha+\mu\right]e^{-\frac12 c \left(\frac{x}{t^\alpha}\right)^2},\n\\
W(x,t)&=& t^\mu e^{-\frac12 c \left(\frac{x}{t^\alpha}\right)^2}.\n
\eea

 %--------------- Ex. 2 ---------
 \subsubsection*{$\bullet$~{\bf Example 2 : $\sigma=\beta z$}}
\medskip
 
 If we match Eq.\,(\ref{ODE2}) with the equation in $\S 2.1.2.69$ of \cite{PZ},
 \[
 zy''+azy'+ay=0,~~~ a>0, ~ z\in [0,\infty),
 \]
 which has a particular solution
 \[
 y_0=z\,e^{-az},
 \]
then we get a CDR system
\bea
C(x,t)&=&t^{\alpha-1}\left((\alpha-\beta a)\,\frac{x}{t^\alpha}+\beta\right),\n\\
D(x,t)&=&\beta\, t^{\alpha-1}\,x,,~\n\\
R(x,t)&=&(\alpha+\mu)\,x\,t^{\mu-\alpha -1}\,e^{- a\frac{x}{t^\alpha}},\n\\
W(x,t)&=& x\,t^{\mu-\alpha }\,e^{- a\frac{x}{t^\alpha}}.\n
\eea

In Fig.\,1 we present the plots of these functions for a set of parameters with three different times.

%-------   Ex. 3-----
 \subsubsection*{$\bullet$~{\bf Example 3 : $\sigma(z)=-\frac12(\alpha+\mu)z^2$}}
\medskip

With
\bea
\sigma(z)&=&-\frac12(\alpha+\mu)z^2,~~~\tau(z)=-\mu z,\n\\
\rho(z)&=&\frac12 (\alpha+\mu) z^2 \left(g^2(z)+g'(z)\right]\,y(z),  
\eea
where $g(z)$ is some function of $z$,
Eq.\,(\ref{ODE2}) becomes
\be
y''-(g^2+g')y=0.
\ee
This equation admits a particular solution
\be
y_0(z)=e^{\int^z g(z) dz}.
\ee

One can choose $g(z)$ such that $y(z)\to 0$ fast enough so that the boundary terms go to zero.
This give us  an exactly solvable CDR system with
\bea
C(x,t)&=&-\mu\,\frac{x}{t};\n\\
D(x,t)&=&-\frac12(\alpha +\mu)\,\frac{x^2}{t},\label{ex4}\\
R(x,t)&=&\frac12 (\alpha+\mu) z^2 \left(g^2(z)+g'(z)\right]\exp\left(\int^z g(z) dz\right),\n\\
W(x,t)&=& t^\mu \exp\left(\int^z g(z) dz\right).\n
\eea

%-----------------  Nonlinear cases -------

\section{Nonlinear cases}

One can obtain nonlinear CDR systems by matching Eq.\,(\ref{ODE2}) with some nonlinear ODEs. We will only illustrate this by a simple example.

%-----    Ex. 4 ---

\subsubsection*{$\bullet$ {\bf Example 4}}

The simple nonlinear ODE
\be
y''-\lambda y^3=0, ~~~\lambda >0,~~ z\in [0,\infty)
\label{nonlinear}
\ee
has solution
\[
y=\left[\sqrt{\frac{\lambda}{2}}\,z+C\right]^{-1}, ~~~~~ C>0.
\]
We can match Eq.\,(\ref{ODE2}) with (\ref{nonlinear}) with, say $\sigma=1$ and $\tau=$constant. 
This leads to
\be
\rho(z)=\mu y+ (\alpha z-\tau)\sqrt{\frac{\lambda}{2}}\,y^2 -\lambda\,y^3.
\ee
 Equation of continuity (\ref{EOCa}), or (\ref{EOCa2}), requires that $\alpha=0$, and thus $\delta=\gamma=-1$. The corresponding nonlinear CDR system is defined by
\bea
C(x,t)&=&\frac{\tau}{t};\n\\
D(x,t)&=&t^{-1},\label{ex-NL}\\
R(x,t)&=&t^{\mu-1}\,\left[\mu y -\tau \sqrt{\frac{\lambda}{2}}\, y^2-\lambda y^3\right]\n,
\eea
with solution
\be
W(x,t)= t^\mu \left[\sqrt{\frac{\lambda}{2}}\,x+C\right]^{-1}, ~~~~~ C>0.
\ee

We show in Fig.\,2 the graphs of $R(x,t)$ and $W(x,t)$ for a set of parameters with three different times.  The functions $C(x,t)$ and $D(x,t)$ are just horizontal lines for fixed time $t$.

% -----      Equivalent system --

\section{Equivalent systems}

It is noted that solutions $W(x,t)$ of some CDR systems are the same as the solutions of some of the Reaction-Diffusion systems given in \cite{HL}, with the same exponents $\alpha$ and $\mu$.  For instance, the solution of Example 1  in this work is the  same as that in Section 6.1 of \cite{HL}.  This prompted us to consider the relation between different CDR systems having the same solution $W(x,t)$ (thus with the same exponents $\alpha$ and $\mu$), but with different diffusion, convention, and reaction terms.  We shall call these systems ``equivalent systems".

Consider the following two CDR systems
\bea
\sigma y''+(\sigma' +\alpha\,z-\tau )\,y' -(\tau'+\mu)\, y+\rho=0,\n\\
\tilde\sigma y''+(\tilde\sigma' +\alpha\,z-\tilde\tau )\,y' -(\tilde\tau'+\mu)\, y+\tilde\rho=0.
\eea

Let these two equations admit the the same solution $y(z)$, then one has
\bea
\frac{d}{dz}\left[(\tilde\sigma-\sigma)y'\right]-\frac{d}{dz}\left[(\tilde\tau-\tau)y\right]+\left(\tilde\rho-\rho\right)=0.
\label{ES}
\eea

Now suppose we are given the solution of the CDR system given by $\{\sigma, \tau, \rho\}$, which includes the Fokker-Planck and reaction-diffusion equations as special cases, and we want to construct an equivalent system $\{\tilde\sigma, \tilde\tau, \tilde\rho\}$. This can be done by specifying two of three functions of 
$\{\tilde\sigma, \tilde\tau, \tilde\rho\}$, and determine the third one from Eq.\,(\ref{ES}).  An exactly solvable new CDR system is then obtained, provided that Eq.\,(\ref{ES}) can be exactly solved.

This construction respects the the equation of continuity (\ref{EOCa}), since Eq.\,(\ref{ES})  implies the identity
\bea
&&\int_{\cal D}\, \rho(z)\,dz + \Delta \left(\sigma\frac{dy}{dz}-\tau y\right)_{\partial {\cal D}}\n\\
=&&\int_{\cal D}\, \tilde\rho(z)\,dz + \Delta \left(\tilde\sigma\frac{dy}{dz}-\tilde\tau y\right)_{\partial {\cal D}}.
\eea

Suppose $\tilde\sigma=\sigma$ and $\tilde \tau$ are given, then the corresponding $\tilde\rho$ of an  equivalent CDR will determined by
\be
\rho+\frac{d}{dz}\left[(\tilde\tau-\tau)y\right].
\label{transf}
\ee

%-----    Ex. 5 ---

\subsubsection*{$\bullet$ {\bf Example 5}}

Example 6.1 of \cite{HL} is a reaction-diffusion equation defined by (in the notations of this work)
\bea
&&\sigma=1, ~~\tau=0,~~y_0=e^{-\frac12 c z^2} \n\\
&&\rho=[c(\alpha-c)z^2+(\mu +c)] y_0.
\eea

If one wants to construct an equivalent CDR system, say  with $\tilde\sigma=\sigma=1, \tilde\tau=\beta z$ for some constant $\beta$, then the $\tilde\rho$ is $\tilde\rho=[c(\alpha-\beta-c)z^2+(\mu +\beta +c)] y_0$.  Thus we have infinitely many equivalent CDR systems as $\beta$ is arbitrary.   Example\, 1 given in this work corresponds to the case with $\beta=\alpha$.

If one chooses $\tilde\sigma=1$ and $\tilde\tau=\beta$, then $\tilde\rho=[c(\alpha-c)z^2-\beta c z+(\mu +c)] y_0$. This can be obtained by Eq.\,(\ref{transf}) from the systems with  $\tau=0$ and $\tau=\beta z$.

%---   Ex. 6 ---
\subsubsection*{$\bullet$ {\bf Example 6}}

We construct a reaction-diffusion equation which is equivalent to the nonlinear CDR system in Example 4 (note that there $\alpha=0$ to satisfies the equation of continuity) .

Setting $\tilde\sigma=1$ and $\tilde\tau=0$, the required $\tilde\rho$ is $\tilde\rho=\mu\, y-\lambda\,y^3$.

%----    Summary ----
\section{Summary}

We have considered  solvability of the convection-diffusion-reaction
equation with both space- and time-dependent convection, diffusion and reaction terms
by means of the similarity method.  There are two types of scaling behaviours of the CDR equation, relating to whether particle number is conserved or not. By introducing the
similarity variable, the convection-diffusion-reaction equation is reduced to an
ordinary differential equation. The reduced ordinary differential equations, namely, Eqs.\,(\ref{ODE2}) and (\ref{ODE4}),  are quite simple in their functional forms. Particularly, Eq.\,(\ref{ODE4}), which  corresponds to the particle conserving cases,  is integrable and its solution can be given in closed form. By matching these
two ordinary differential equations with some known exactly solvable equations, one can obtain corresponding exactly solvable CDR  systems.  Some representative examples of exactly solvable systems were presented.
Finally, we have briefly discussed how an equivalent CDR system can be constructed which admits the same similarity solution of another CDR system. 

% -----------------------------------------------------------------------------------------------------------------------------------------------------------------------------------------------

\acknowledgments

The work is supported in part by the Ministry of Science and Technology (MoST)
of the Republic of China under Grant MOST 106-2112-M-032-007.

%---------------------------------

%\bibliographystyle{plain}

%\newpage

%---------------  Figures ------

%---    Fig. 1 ---
\begin{figure}[ht] \centering
\includegraphics*[width=8cm,height=6cm]{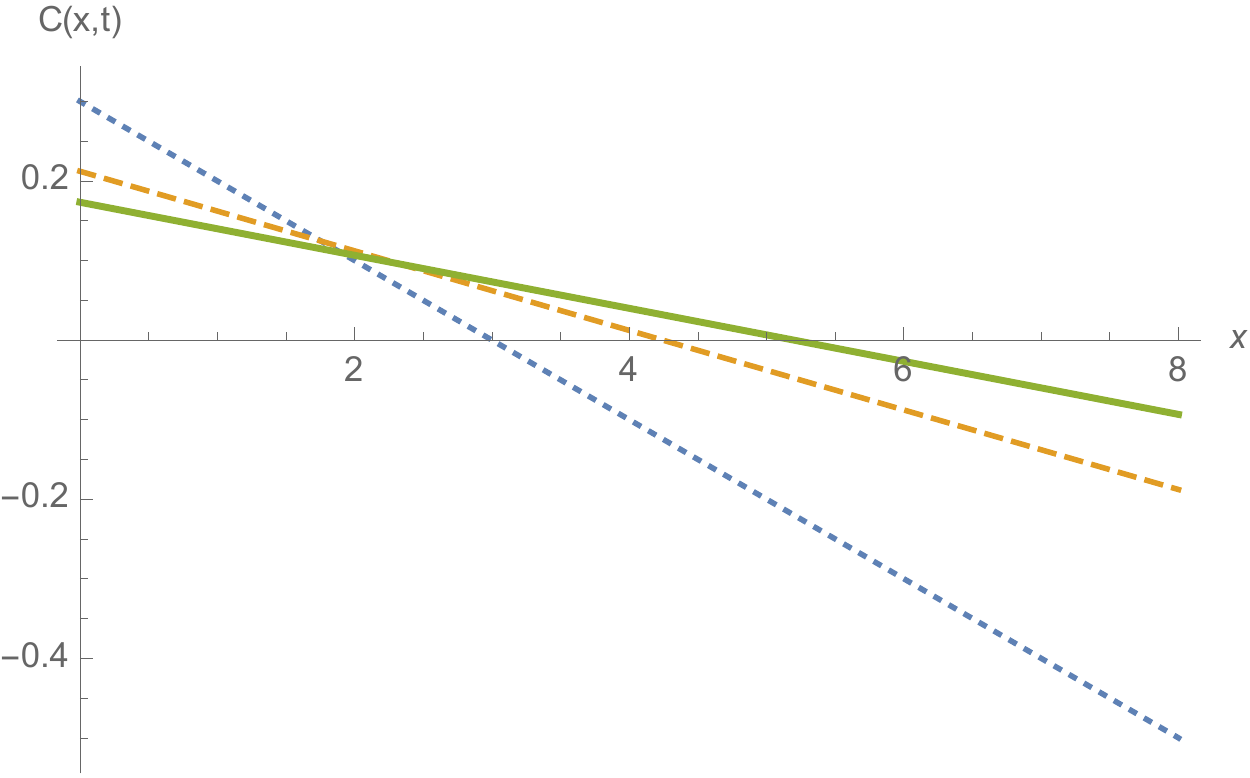}\hspace{1cm}
\includegraphics*[width=8cm,height=6cm]{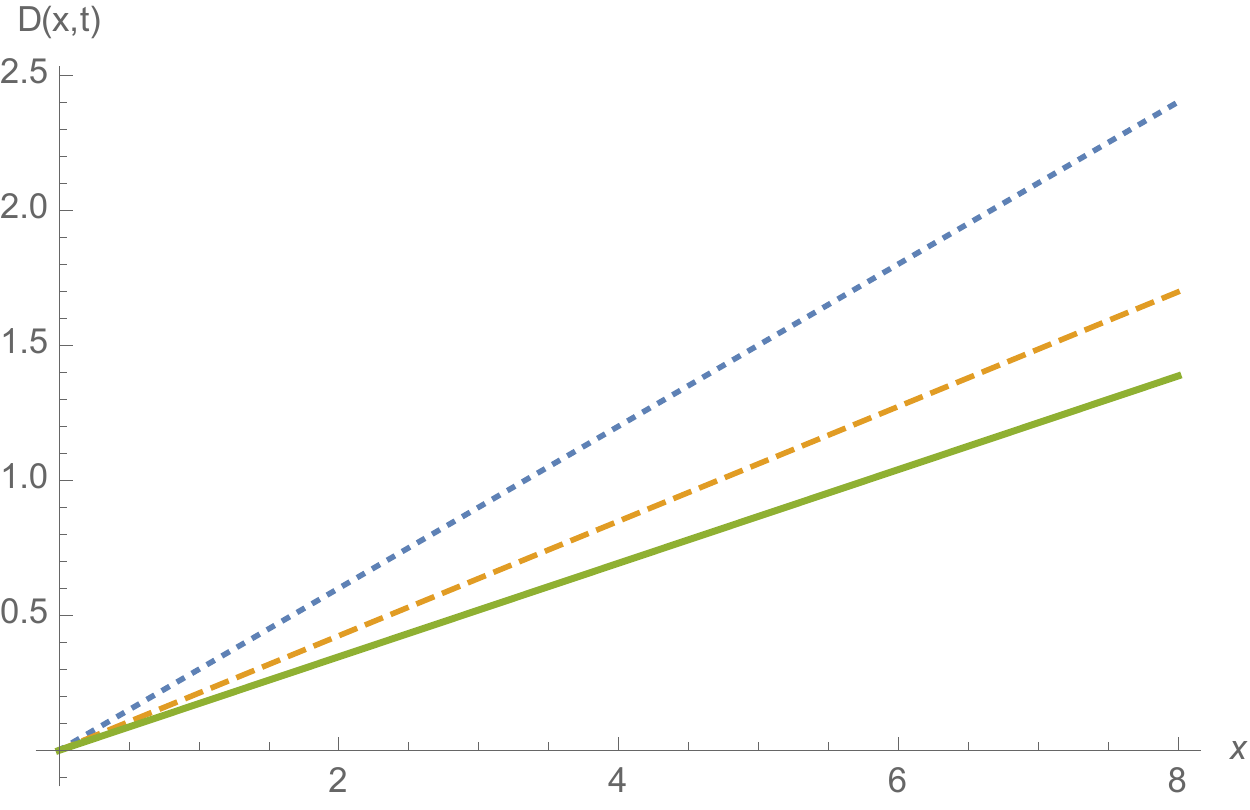}\\
\includegraphics*[width=8cm,height=6cm]{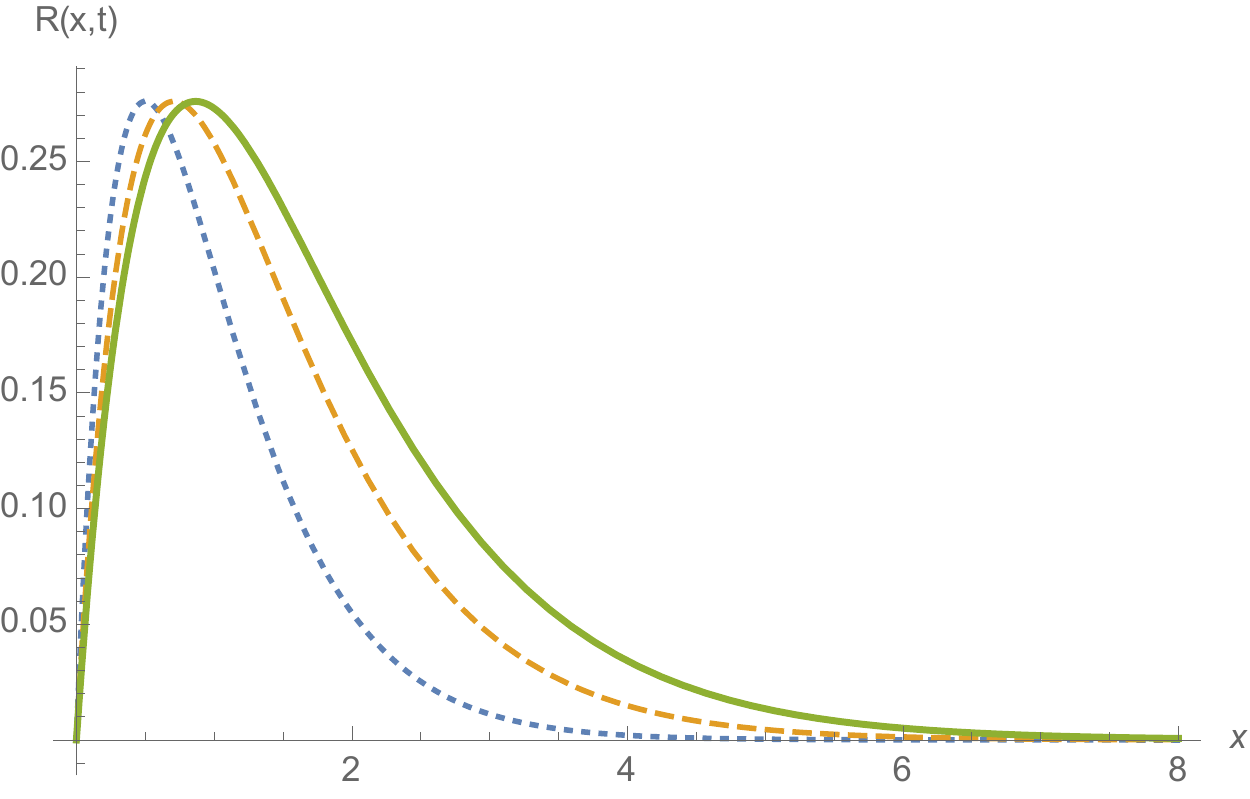}\hspace{1cm}
\includegraphics*[width=8cm,height=6cm]{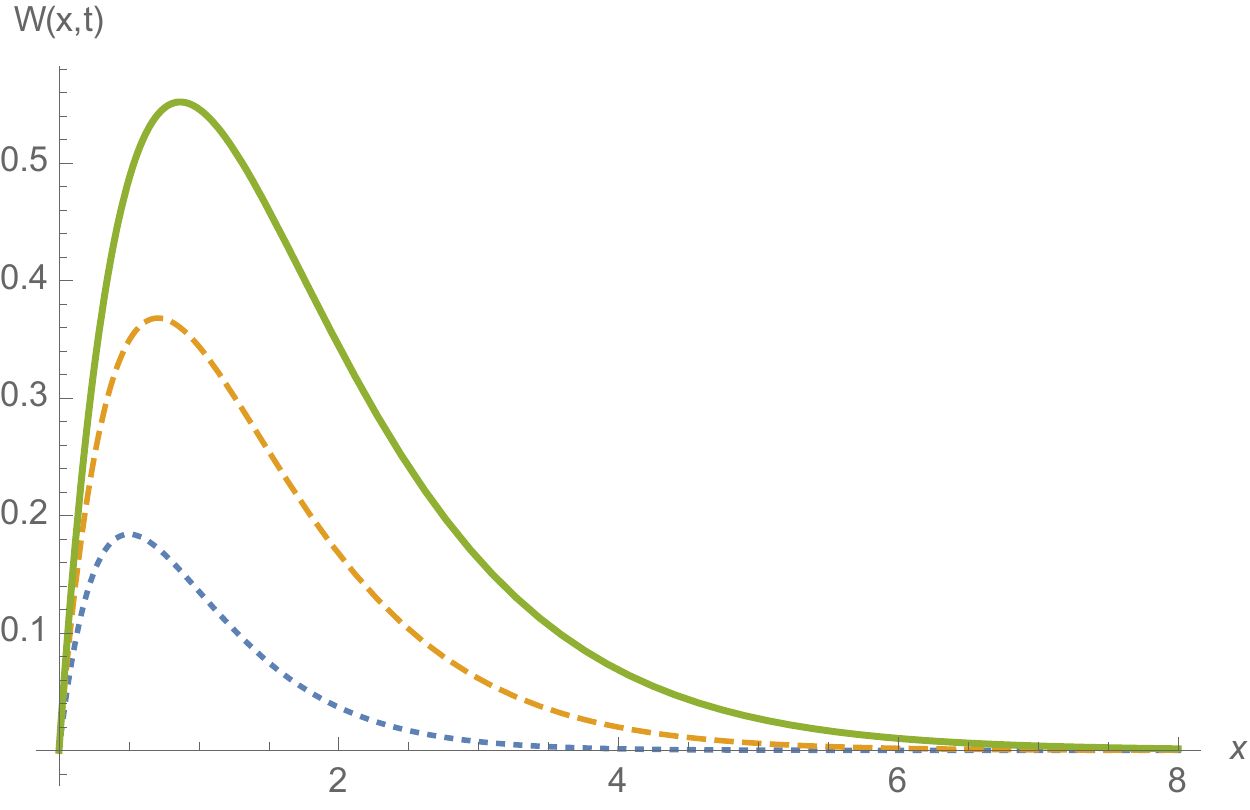}
\caption{Plot of $C(x,t),D(x,t), R(x,t)$ and $W(x,t)$ for Example  2  with $\alpha = 0.5,  \beta= 0.3, \mu = 1, a=2 $, and time $t = 1$ (dotted), $t = 2$ (dashed), $t = 3$ (solid). }
\label{Fig1}
\end{figure}
%------------------

%---    Fig. 2 ---
\begin{figure}[ht] \centering
\includegraphics*[width=8cm,height=6cm]{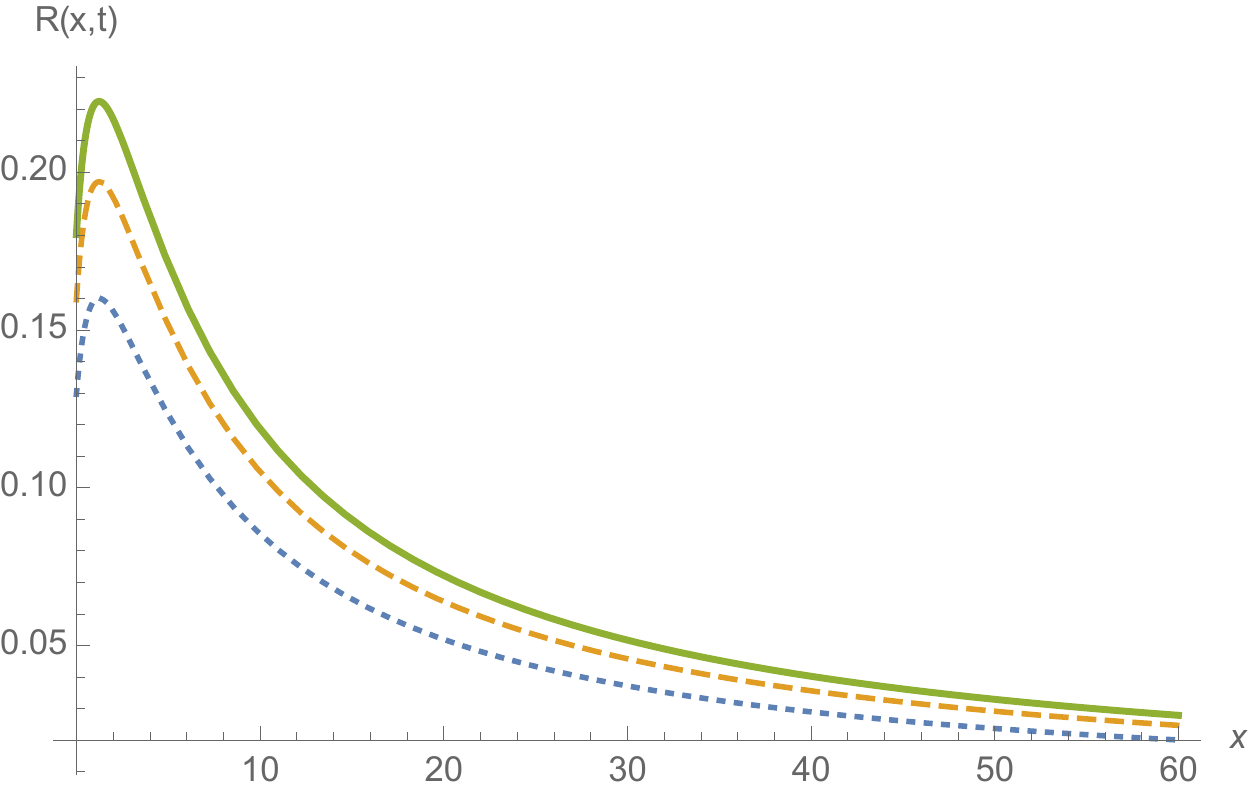}\hspace{1cm}
\includegraphics*[width=8cm,height=6cm]{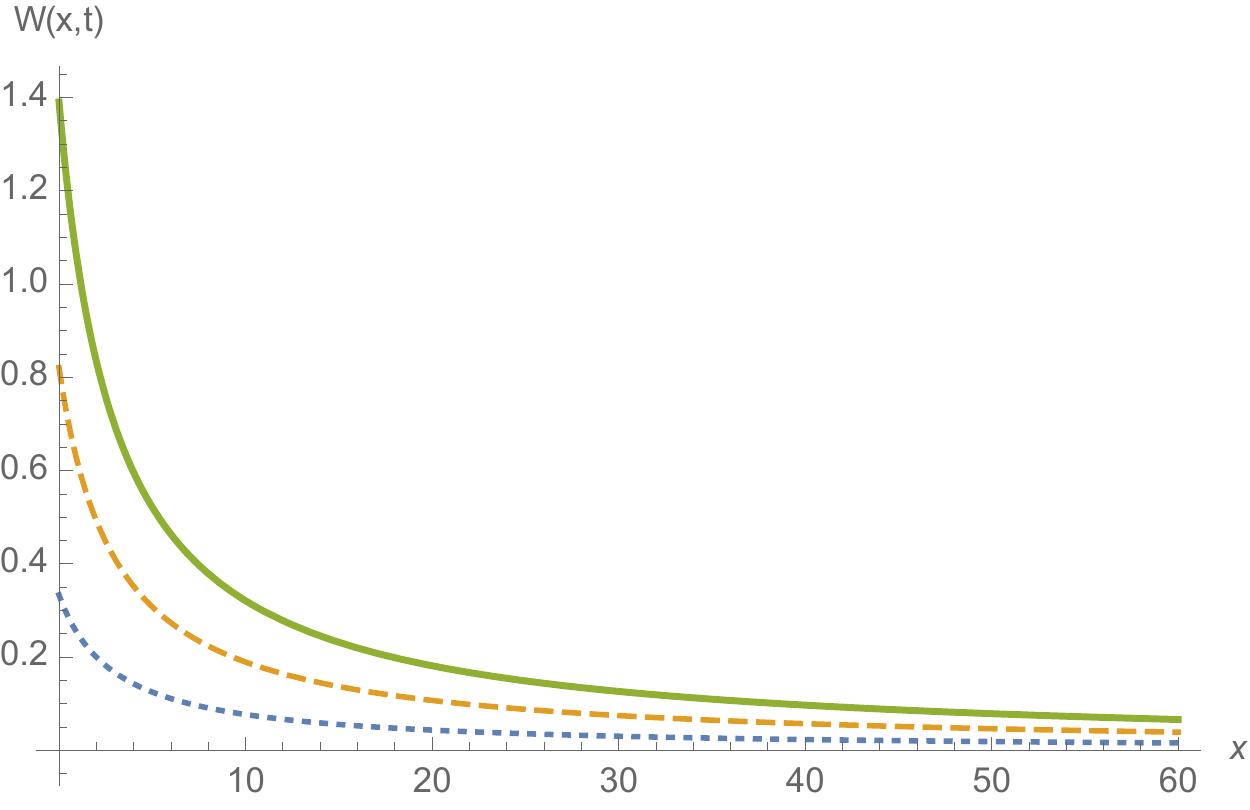}
\caption{Plot of $R(x,t)$ and $ W(x,t)$ for Example 4 with $\alpha = 0,
\mu = 1.3,  \lambda=2, \tau=2.4,C = 3$ and time $t = 1$ (dotted), $t = 2$ (dashed),
$t = 3$ (solid).} \label{Fig2}
\end{figure}

\end{document}